\documentclass[showpacs, twocolumn]{revtex4-1}
\usepackage{graphicx}
\usepackage{amsmath}
\usepackage{appendix}
\begin{document}

\title{Quantum liquid droplets in Bose mixtures with weak disorder}

\author{Karima Abbas$^{1,2}$ and Abdel\^{a}ali Boudjem\^{a}a$^{1,2}$}
\affiliation{$^1$ Department of Physics, Faculty of Exact Sciences and Informatics, Hassiba Benbouali University of Chlef, P.O. Box 78, 02000, Ouled-Fares, Chlef, Algeria. \\
$^2$Laboratory of Mechanics and Energy, Hassiba Benbouali University of Chlef, P.O. Box 78, 02000, Ouled-Fares, Chlef, Algeria.}
\email {a.boudjemaa@univ-chlef.dz}

\date{\today}

\begin{abstract}

We study the properties of self-bound liquid droplets of three-dimensional Bose mixtures in a weak random potential with Gaussian correlation function at both zero and finite temperatures. 
Using the Bogoliubov theory, we derive useful  formulas for the ground-state energy, the equilibrium density, the depletion,  and the anomalous density of the droplet. 
The quantum fluctuation induced by the disorder known as the glassy fraction is also systematically computed.  
At finite temperature, we calculate the free energy, the thermal equilibrium density, and the critical temperature in terms of the disorder parameters. 
We show that when the strength and the correlation length of the disorder potential exceed a certain critical value, the droplet evaporates and is eventually entirely destroyed.
We calculate the density profiles of this exotic state by means of numerical simulations of the corresponding generalized disorder Gross-Pitaevskii equation.
Our predictions reveal that as the strength of the disorder gets larger, the atomic density varies rapidly in the plateau region.
We point out in addition that the peculiar interplay of the disorder and the repulsive Lee-Huang-Yang corrections play a pivotal role in the collective modes of the self-bound droplet.

\end{abstract}

\maketitle

\section{Introduction} \label{Intro}

Ultracold Bose-Bose mixtures of atomic gases in disordered media have attracted tremendous attention due to the 
high degree of control over  interspecies interactions and disorder effects (see e.g.\cite{Boudj1, Boudj2,Boudj3}). 
An eminent property of such binary Bose-Einstein condensates (BECs) is the possibility to form quantum self-bound droplets, originating from the
competition of mean-field attraction and beyond-mean-field repulsion  provided by the Lee-Huang-Yang (LHY) corrections \cite{Petrov,Cab,Sem,Err}. 
The same exquisite stabilization mechanism leads to the formation of  self-bound droplets in single and binary dipolar BECs, where the balance between the mean-field energy  
associated with short and long-range interactions and the LHY corrections arrests the dipolar instability at high density \cite{Pfau,Pfau1,Chom,Boudj4,Boudj5,Boudj6}.
Quantum liquid droplets have been widely explored in various contexts (see for review \cite {Luo,Pfau2,Guo} and references therein).

The aim of this paper is to present a comprehensive understanding of the properties of self-bound liquid droplets in a weak three-dimensional (3D) Gaussian-correlated disorder. 
This latter is defined by both the strength and the correlation length of the disorder (see below) which enables us to well control the interplay between
the disorder correlation length, the LHY fluctuations and the  interspecies interaction strength. 
Quite recently, two-dimensional droplets in binary BECs subjected to a random repulsive potential have been studied numerically in \cite{Sahu}.
The dirty self-bound liquids can be regarded as a feasible simulator to analyze a plethora of novel quantum phenomena.
We look in particular at how the peculiar competition between the interspecies interactions, the LHY quantum fluctuations, and the disorder potential 
affect the formation and the stability of the self-bound droplet. The role of the LHY corrections in dirty dipolar and nondipolar BECs has been discussed in Refs.\cite{Boudj44,Nagler}.

Here we  derive useful analytical expressions for relevant physical quantities such as the equation of state, the equilibrium density, the glassy fraction, the depletion 
and the anomalous density of the droplet at both zero and finite temperatures within the framework of the Bogoliubov theory. 
The approach is valid provided that the disorder-induced depletion is sufficiently weak. 
The presence of the disorder fluctuations furnishes an additional term which competes with the mean-field and beyond-mean-field LHY nonlinearities leading to modify
the properties of the droplet.
Our work is also surmised on the observation that at increasing disorder strength and decreasing correlation length, the self-bound droplet splits into mini droplets trapped in
small wells of the disorder potential due to the destruction of the coherence.
We find that the equilibrium density decays with the disorder parameters.
Consequently, there exists a certain critical value of the strength and the correlation length of the disorder potential above which the droplet evaporates and eventually completely disappears.
Such a critical strength  strongly depends on the interspecies interactions and on the disorder correlation length.
We then extend our study to the finite temperature case. 
It is shown that the disorder substantially influences the thermal equilibrium and the critical temperature of the self-bound liquid droplet.

On the other hand, we derive a generalized disorder Gross-Pitaevskii equation (GPE) employing the local density approximation.
The numerical simulations of this equation reveals that the density follows the modulations of the disorder in the plateau region. 
Nevertheless, the action of the disorder is irrelevant near the edges of the droplet.
The collective modes of the droplet are analyzed by means of a variational Gaussian ansatz.
We emphasize that the disorder lowers the frequency of the breathing oscillations.

The rest of this paper is organized as follows. 
Section \ref{BHM} introduces the Bogoliubov-Huang-Meng model for dirty Bose-Bose mixtures.
We then use the developed model to derive analytical expressions for various physical quantities, including the equation of state, the glassy fraction, and the normal and anomalous densities.
Section \ref{SBD} is concerned with the ground-state properties of a disordered self-bound droplet at both zero and finite temperatures. 
We look in particular at how the disorder potential affects the existence and the formation of the liquid droplet.
The equilibrium density, the critical disorder strength, and the critical temperature are accurately calculated in terms of the system parameters.
Section \ref{GDGPE} is dedicated to the numerical analysis of the generalized disorder GPE which  successfully describes density profiles of the droplet in the equilibrium state.
In Sec.\ref{modes} we calculate the collective modes of the disordered droplet utilizing a variational method.   
Finally, in Sec.\ref{Conc}, we present a summary of our conclusions.

\section{Model} \label{BHM}

We consider a weakly interacting homogeneous Bose mixture with equal masses $m_1 = m_2 = m$, subjected to a 3D weak disorder potential $U({\bf r})$. 
The disorder potential is assumed to have vanishing ensemble averages $\langle U({\bf r})\rangle=0$
and a finite correlation of the form $\langle U({\bf r}) U({\bf r'})\rangle=R ({\bf r}-{\bf r'})$.
The Hamiltonian including all point-like interactions can be written in terms of the creation and annihilation operators $\hat a^\dagger_{j,\bf k}$ and $\hat a_{j,\bf k}$,
where the subscript $j = \{1, 2\}$ refers to the $j^{th}$ component of the mixture, as:
\begin{align}\label{ham1}
&\hat H\!\!=\!\!\sum_{j,\bf k}\! E_k\hat a^\dagger_{j,\bf k}\hat a_{j,\bf k}\! 
+\!\frac{1}{V}\!\!\sum_{j,\bf k,\bf p} \! U_{\bf k\!-\!\bf p} \hat a^\dagger_{j,\bf k} \hat a_{j,\bf p}  \\
&+\!\frac{g_j}{2V}\!\!\sum_{j,\bf k,\bf p,\bf q}\!\!\hat a^\dagger_{j,\bf k} \hat a^\dagger_{j,\bf p}\hat a_{j,\bf p\!-\!\bf q}\hat a_{j,\bf k\!+\!\bf q}  \nonumber\\
&+\!\frac{g_{12}}{V}\!\!\sum_{\bf k,\bf p,\bf q}\!\!\hat a^\dagger_{1,\bf k} \hat a^\dagger_{2,\bf p}\hat a_{2,\bf p\!-\!\bf q}\hat a_{1,\bf k\!+\!\bf q}, \nonumber
\end{align}
where $E_k=\hbar^2k^2/2m$, $V$ is a quantization volume, and  $U_{\bf k}$ is the Fourier transform of the external random potential $U({\bf r})$.
In the Hamiltonian (\ref{ham1}), we have  introduced the coupling constants for the intraspecies interactions $g_j=4\pi \hbar^2a_j/m$ 
as well as for the interspecies interaction $g_{12}=g_{21}= 4\pi \hbar^2a_{12}/m$ with 
$a_j$ and $a_{12}$ being the intraspecies and the interspecies scattering lengths, respectively.

The elementary excitations and the ground-state energy of the mixture are obtained by applying the Bogoliubov prescription 
which consists of replacing the operators $\hat a_{j,0}$ and $\hat a^\dagger_{j,0}$  by  a $c$-number, i.e., $\hat a_{j,0}= \hat a^\dagger_{j,0}=\sqrt{N_{jc}}$, where $N_{jc}$ is 
the number of condensed particles.
In the resulting equation, we ignore higher-order fluctuations and keep only terms in $\hat a^\dagger_{{j,\bf k} \neq0}$,  $\hat a_{{j,\bf k} \neq0} $ 
up to the second-order in the coupling constants.
In the context of disordered BECs, the Bogoliubov theory suggests that for weak enough disorder,  disorder fluctuations decouple in the lowest order \cite{HM,Gior}. 

To diagonalize the Hamiltonian (\ref{ham1}), we introduce the Bogoliubov-Huang-Meng transformation \cite{HM}:
\begin{subequations}\label {T:DH}
\begin{align}
\hat a_{1,\bf k}&= u_{+k} \hat b_{1,\bf k}-v_{+k} \hat b^\dagger_{1,-\bf k}-\beta_{1,\bf k}, \\
 \hat a_{2,\bf k}&=u_{-k} \hat b_{2,\bf k}-v_{-k} \hat b^\dagger_{2,-\bf k}-\beta_{2,\bf k}, 
\end{align}
\end{subequations}
where $\hat b^\dagger_{j,\bf k}$ and $\hat b_{j,\bf k}$ are  operators of elementary excitations obeying the usual Bose commutation relations,
the quasiparticle amplitudes are given by:
\begin{equation}  \label{BogAmp}
u_{\pm,k} =\frac{1}{2}\left (\sqrt{ \frac{\varepsilon_{\pm,k}} {E_k}} + \sqrt{\frac{E_k}{\varepsilon_{\pm,k}}} \right), \;\; v_{\pm,k}=u_{\pm,k}-\sqrt{\frac{E_k} {\varepsilon_{\pm,k}}},
\end{equation}
 and the disorder translations $\beta_{1,k}$ are defined by the equations
\begin{align}  \label{Beta}
&\beta_{1,\bf k}=\sqrt{\frac{n_1}{V}}  \frac{|u_{+,k}-v_{+,k}|^2} {\varepsilon_{+,k}} U_{\bf k}, \\
&\beta_{2,\bf k}=\sqrt{\frac{n_2}{V}}  \frac{|u_{-,k}-v_{-,k}|^2} {\varepsilon_{-,k}} U_{\bf k}, \nonumber
\end{align}
where  the excitation spectrum energies $\varepsilon_{\pm,k}$ are defined below.

In this work we consider the case of a symmetric mixture with $n_1=n_2=n/2$ and $g_1=g_2=g$. 
The Bogoliubov excitation energies, which turn out independent on disorder for the symmetric mixture studied here, reads 
$\varepsilon_{k\pm}=\sqrt{E_k^2 +2 E_k n \delta g_{\pm}}$, where  $\delta g_{\pm}=g(1\pm g_{12}/g)$ \cite{Boudj7}.

Performing the diagonalization via (\ref{BogAmp}) and the average over the disorder, the Hamiltonian (\ref{ham1}) transforms into:
\begin{equation}  \label{ham2}
\hat H\!\!= E+ \sum_{{\bf k}\neq0} \left(\varepsilon_{k,+} \hat b^\dagger_{1\bf k} \hat b_{1\bf k}+\varepsilon_{k,-} \hat b^\dagger_{2\bf k} \hat b_{2\bf k}\right), 
\end{equation}
where the ground-state energy of the system including the disorder corrections is given by
\begin{align}\label{genergy} 
\frac{E}{V}&=\sum\limits_{\pm} \bigg[ \frac{1}{2} n^2 \delta g_{\pm} - \frac{n}{2\pi^2} \int_0^{\infty} dk\,k^2  R_k \frac{ E_k}{\varepsilon_{k \pm}^2}\\
&+\frac{1}{4\pi^2} \int_0^{\infty} dk\, k^2 \left(\varepsilon_{k \pm} -2E_k- n \delta g_{\pm}+ \frac{n^2 \delta g_{\pm}^2}{2E_k}\right)\bigg], \nonumber
\end{align}
The leading term is the mean-field energy. The subleading term gives the correction to the ground-state energy due to the external random potential.
The last term accounts for the regularized ground-state energy owing to the LHY quantum corrections.

The noncondensed $\tilde n =\sum\limits_{j\bf k} \langle \hat a^\dagger_{j,\bf k} \hat a_{j,\bf k}\rangle$ and anomalous 
$\tilde m=\sum\limits_{j\bf k} \langle \hat a_{j,\bf k} \hat a_{j,\bf k}\rangle$ densities read as \cite{Boudj18,Boudj21}
\begin{equation}\label {nor}
\tilde n_{\pm}=\frac{1}{4\pi^2} \int_0^{\infty} dk\, k^2 \left[\frac{E_k+ n \delta g_{\pm}} {\varepsilon_{k \pm}} \sqrt{I_{k \pm}}-1\right]+n_{R\pm},
\end{equation}
and
\begin{equation}\label {anom}
\tilde m_{\pm}=-\frac{1}{4\pi^2} \int_0^{\infty} dk\, k^2 \frac{ n \delta g_{\pm} } {\varepsilon_{k \pm}} \sqrt{I_{k\pm}}+n_{R\pm},
\end{equation}
where $I_{k\pm} =\text {coth} ^2\left(\varepsilon_{k\pm}/2T\right)$ with $T$ being the temperature \cite{Boudj7,Boudj18,Boudj21}.
The Boltzmann constant set $k_B=1$ throughout the manuscript.\\
The last term in Eqs.(\ref{nor}) and (\ref{anom}) can be evaluated through Eq.(\ref{Beta}):
\begin{equation} \label{glass}
n_{R\pm}=\frac{ 1}{ V}\sum_{\bf k} \langle |\beta_{{\bf k}\pm}|^2 \rangle=\frac{n}{2\pi^2} \int_0^{\infty} dk\,k^2\frac{ E_k^2}{ \varepsilon_{k\pm}^4} R_k,
\end{equation}
which  accounts for the disorder fluctuations known also as {\it glassy fraction}.

In what follows we consider a disorder potential with Gaussian autocorrelation function which is very popular in ultracold atom experiments. 
The corresponding Fourier transform  has the form \cite{Kob,Boudj8}:
\begin{equation} \label{CorrS}
R_k = R_0 \, e^{-\sigma^2k^2/2},
\end{equation} 
where $R_0$ with dimension (energy) $^2$ $\times$ (length)$^3$ and $\sigma$ characterize respectively, the strength and the correlation length of the disorder.
For $\sigma \rightarrow 0$, the Gaussian potential reduces to the uncorrelated $\delta$-random potential.
This simple model is useful in three respects. Firstly, it can be realized either with speckle potentials or with Gaussian impurity disorders \cite{LCLBA}.
Secondly, it is analytically tractable, and therefore provides a test for the numerical simulation.
Thirdly, the energy shift due to the Gaussian-correlated disorder (\ref{CorrS}) does not require any regularization (i.e. safe from ultraviolet divergence)
unlike the $\delta$-correlated random potential made of a random series of $\delta$ peaks as we shall see in Eq.(\ref{genergy1}).
Note that our formalism can be applied for any correlated and uncorrelated disorder potentials.

\section{Self-bound droplets} \label{SBD}

In this section,  we apply the developed approach to investigate the formation and the stabilization of self-bound droplets  in the presence of 3D disorder at 
both zero and finite temperatures.

\subsection{Zero temperature} \label{T0}

Substituting Eq.(\ref{CorrS}) into Eq.(\ref{genergy}) and integrating over $k$ we obtain the total energy. Normalizing the resulting energy and the density to their equilibrium
values obtained within the theory of Petrov for clean droplets, namely: $n^{(0)}=25 \pi (\delta a_+/a)^{2}/(16384 a^{3})$ and 
$\vert E_{0}\vert/N =25\pi^{2}\hbar^{2}\vert \delta a_+/a\vert^{3}/ (49152ma^{2})$ \cite {Petrov,Ota}, we then find:
\begin{align} \label{genergy1} 
\frac{E}{|E_{0}|}&= -3\left(\frac{n}{n^{(0)}}\right)+\frac{1}{2\sqrt{2}}\left(\frac{n}{n^{(0)}}\right)^{3/2}\sum_{\pm} \left(\frac{\delta a_{\pm}}{a}\right)^{5/2} \\
&-\sum_{\pm}\frac{R \vert \delta a_+/a\vert}{24\sqrt{2}\pi(\sigma/\xi)} \Bigg[\frac{1}{\sqrt{\pi}}-\frac{\sigma}{\xi_{\pm}} e^{\sigma^2/\xi_{\pm}^2} 
\text{erfc}\left(\frac{\sigma}{\xi_{\pm}}\right) \Bigg], \nonumber
\end{align}
where $\text{erfc} (x)$ is the complementary error function, $R=R_{0} N^2/ (\xi^{3} E_0^2)$, 
and $\xi_{\pm}=\xi/\sqrt{ (n/n^{(0)}) (\delta a_{\pm}/a)}$ with $\xi= \hbar/\sqrt{mgn^{(0)}}$.
Evidently, for $R=0$, Eq.(\ref{genergy1}) reduces to that found for clean droplets \cite{Petrov,Boudj18, Ota,Hu}.
For $\sigma /\xi \rightarrow 0$, the energy correction due to the disorder contribution reduces to the result of $\delta$-correlated disorder potential, namely,
$ R\big(\vert \delta a_+/a\vert/24\sqrt{2}\pi \big) \sqrt{(\delta a_{\pm}/a) (n/n^{(0)}) }$.

In the droplet regime $g > 0$ and $g_{12} < 0$, the ground-state energy (\ref{genergy1}) becomes complex predicting a collapse mean-field solution. 
To cure this issue one may simply set $|\delta g_+|/g \ll 1$ \cite {Petrov}. 
Note that the energy can also be  stabilized by taking into account higher-order quantum corrections \cite{Boudj18,Boudj21,Ota,Hu}.

\begin{figure}
\includegraphics[scale=0.8]{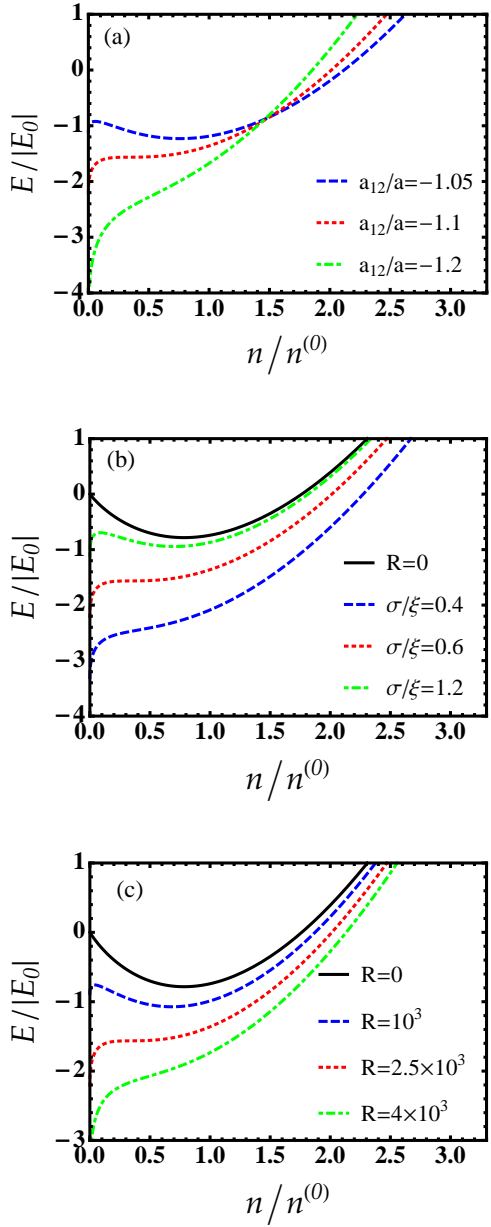}
 \caption{ (a) Ground-state energy from Eq.(\ref{genergy1}) as a function of the density $n$ for various values of the interaction strength $a_{12}/a$ and 
for $R=2.5 \times 10^3$ and $\sigma/\xi=0.63$.
(b) Ground-state energy from Eq.(\ref{genergy1}) as a function of the density $n$ for various values of the disorder correlation length $\sigma/\xi$ and for
$a_{12}/a=-1.1$ and $R=2.5\times 10^3$.
(c) Ground-state energy from Eq.(\ref{genergy1}) as a function of the density $n$ for various values of the disorder strength $R$ and for $a_{12}/a=-1.1$ and $\sigma/\xi=0.63$.}
\label{GSE} 
\end{figure}

In order to prove the relevance of our theory for current experiments, we consider the ${}^{39}$K mixture droplets \cite{Cab}.
The intraspecies scattering length is chosen to $a=71a_{0}$ with $a_0$ being the Bohr radius. 
Note that $a_{12}$ which can be adjusted via the Feshbach resonances is selected in such a way that the droplet phase is reached.
The disorder strength $R_0$ used in the experiment ranges from $ \sim 1.26.10^{-82}$ J$^{2}$.m$^{3}$ to $\sim 1.52.10^{-80}$J$^{2}$.m$^{3}$ (or equivalently
$R$ from 75.34 to 9116.77),  and the correlation length is $\sigma \simeq 0.13 \,\mu$m \cite{Jen}.

In Fig.\ref{GSE}  we show the ground-state energy using different values for the disorder parameters  and interspecies interactions.
One can clearly identify two regions:
In the low density regime where $n \lesssim n^{(0)}$, we see that for fixed values of $R$ and $\sigma/\xi$, the energy decreases 
with the interspecies interactions $|a_{12}/a|$ and then the local minimum of the energy starts to disappear revealing the evaporation of the self-bound droplet (see Fig.\ref{GSE}.(a)). 
This can be attributed to the peculiar competition between the disorder, interaction and quantum fluctuations.
The situation is inverted for $n \gtrsim n^{(0)}$ where the energy increases with $|a_{12}/a|$ in agreement with the case of a clean droplet \cite{Ota, Boudj9}.

Figures \ref{GSE}.(b)  and \ref{GSE}.(c) show that a robust self-bound droplet survives only for large correlation length $\sigma/\xi$ and small disorder strength $R$.
However for small $\sigma/\xi$ and large $R$, the shape of the energy curve does not contain a local minimum causing the formation of an unstable self-bound state.
In such a situation, the droplet most likely segregates into multiple mini droplets similarly to the two-dimensional case \cite{Sahu}.
For large enough $\sigma/\xi$, the energy simplifies to that of  clean droplets ($R=0$) indicating that the disorder effects are not important in this regime (see Fig.\ref{GSE}.(b)).
Once the disorder parameters exceed their critical values (i.e. $R > R_c \simeq 2.5 \times 10^3$ and $\sigma/\xi <\sigma_c/\xi \simeq 0.6$), 
the atoms leave the droplet and accumulate in the depleted region. 
Therefore, the self-bound state loses its intriguing self-evaporation phenomenon and is eventually completely destroyed..

\begin{figure}
\centerline{
\includegraphics[scale=0.8]{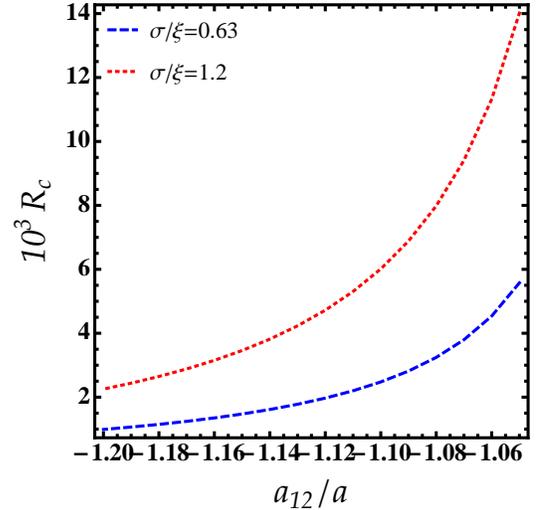}}
 \caption{ Critical disorder strength as a function of $a_{12}/a$ for two different values of $\sigma/\xi$.}
\label{PD} 
\end{figure}

The critical disorder strength $R_c$, for which the local minimum of the energy  disappears, in terms of the interspecies interactions $a_{12}/a$ for two different values of $\sigma/\xi$ 
is captured in Fig.\ref{PD}. We see that $R_c$ increases with $a_{12}/a$ and decays with decreasing $\sigma/\xi$. 

\begin{figure}
\centerline{
\includegraphics[scale=0.8]{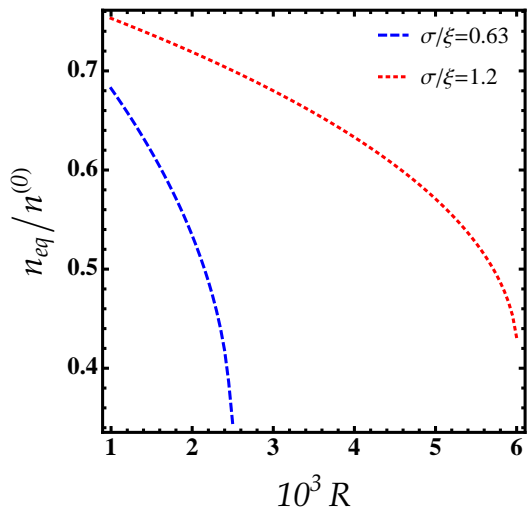}}
 \caption{ Equilibrium density of a dirty droplet $n_{\text{eq}}$ with respect to $n^{(0)}$  as a function of the disorder strength for various values of  $\sigma/\xi$ and $a_{12}/a=-1.1$. }
\label{EquD} 
\end{figure}

The equilibrium density of a dirty droplet $n_{\text{eq}}$ can be obtained by minimizing  the ground-state energy (\ref{genergy1}) with respect to the density. 
Its behavior as a function of the disorder strength for various values of $a_{12}/a$ and $\sigma/\xi$ is depicted in Fig.\ref{EquD}.
Remarkably, the equilibrium density decreases with increasing the disorder strength regardless of the values $a_{12}/a$ and $\sigma/\xi$ 
leading to destabilizing the droplet and eventually destroying it.

\begin{figure}
\includegraphics[scale=0.8]{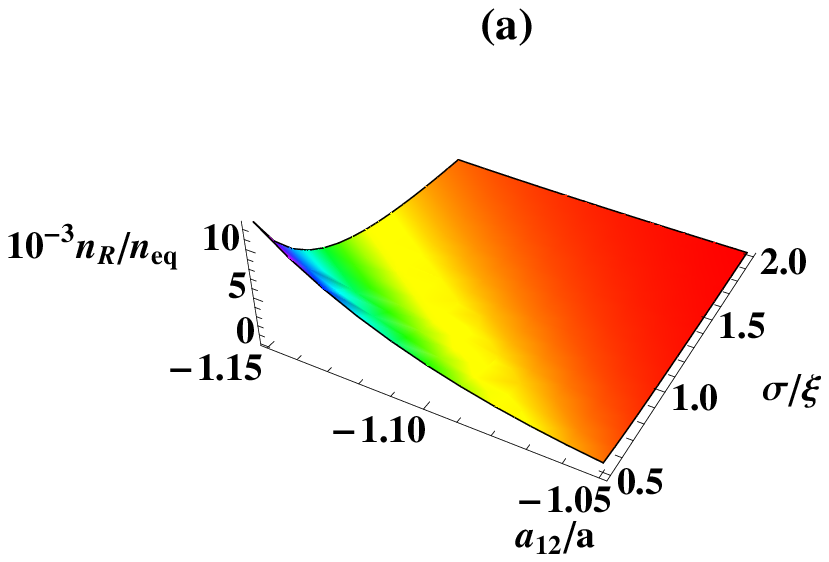}
\includegraphics[scale=0.8]{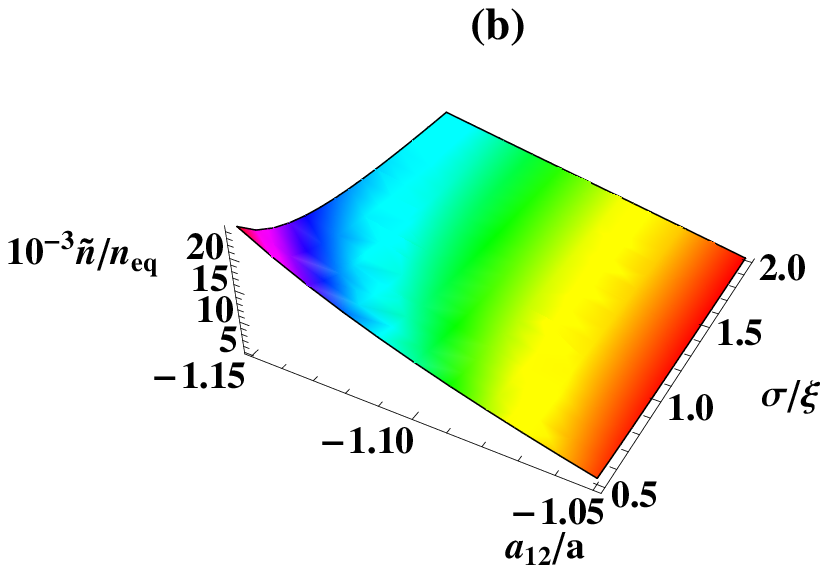}
\includegraphics[scale=0.8]{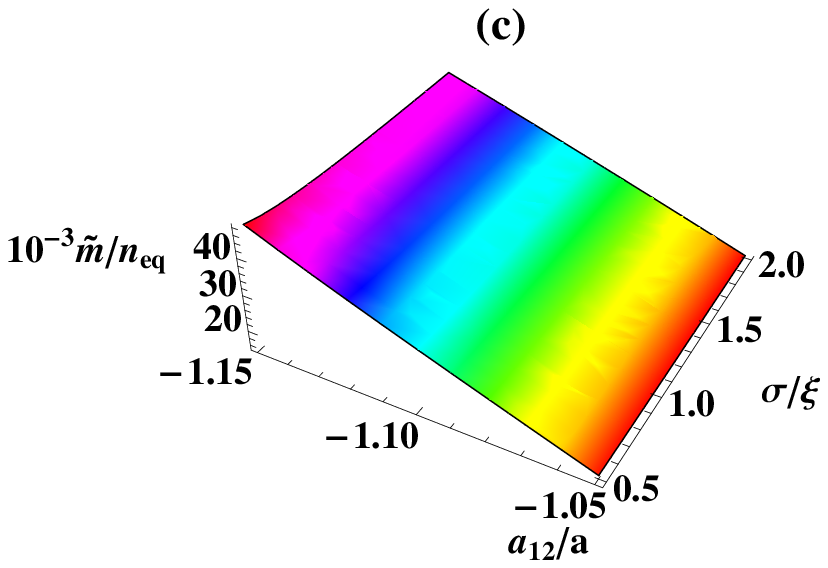}
\caption{ (a) Glassy fraction inside the droplet $n_R$ as a function of $a_{12}/a$ and $\sigma/\xi$ for $R=1000$.
(b) Condensed depletion of the droplet $\tilde n=\tilde n_{0-}+ n_R$  as a function of $a_{12}/a$ and $\sigma/\xi$ for $R=1000$.
(c) Anomalous density of the droplet $\tilde m= \tilde m_{0-}+ n_R$  as a function of $a_{12}/a$ and $\sigma/\xi$ for $R=1000$.}
\label{GL} 
\end{figure}

Introducing the function (\ref{CorrS}) into Eq.(\ref{glass}) and performing the integration over the momentum, we obtain for the glassy fraction inside the droplet
\begin{align}  \label {glass1}
\frac{n_{R}}{n}&=\frac{\sqrt{2}R(\delta a_+/a)^{2}}{144\pi\sqrt{\left(\frac{n}{n^{(0)}}\right)\left(\frac{\delta a_-}{a}\right)}}\\
&\times\Bigg[-\frac{\sigma}{\sqrt{\pi}\xi_{-}}+\left(\frac{1}{2}+\left(\frac{\sigma}{\xi_{-}}\right)^{2}\right) e^{\sigma^2/\xi_{-}^2} \text{erfc}\left(\frac{\sigma}{\xi_{-}}\right)\Bigg].\nonumber
\end{align}
This equation is appealing since it explains the interplay between the disorder potential, the LHY quantum corrections and the attractive interspecies interaction.
For $\sigma \rightarrow 0$, Eq.(\ref{glass1}) reduces to $n_R/n= (\delta a_+/a)^2 R/ \big[144 \pi \sqrt{2( n/n^{(0)}) (\delta a_-/a)}\big]$ which corresponds to the results of
the white-noise disorder potential. For $\sigma \rightarrow 0$ and $a_{12}=0$, one recovers the seminal results of Huang-Meng for a single dirty Bose gas \cite{HM}.
We see from  Fig.\ref{GL}.(a) that in the region $\sigma \rightarrow 0$ and for small $a_{12}/a$, the total glassy fraction $n_R/n$ is significant. 
By increasing the disorder strength or reducing the interspecies interaction, the macroscopic occupation of the ground-state decreases more and more, hence  the fragmented droplets 
in the small wells of the random potential increase even at zero temperature due to the randomness. 
Lifting further the disorder strength, one can expect that the droplet becomes completely depleted, giving rise to destruction of the coherence.
For relatively large $a_{12}/a$ and $\sigma \gg \xi$, ${n_R}$ becomes negligibly small, indicating that the atoms are less localized in such a regime, 
leading to the formation of an extended droplet.
This delocalization can be interpreted as the fact that the disorder potential is screened by both the LHY quantum corrections and the interactions.
This behavior also holds true  in the case of a dirty ordinary BEC (see e.g.\cite{Boudj8, Axel}).

For ${}^{39}$K atoms we have  $a=71a_{0}$ \cite{Cab} and for $a_{12}/a = -1.05$, $R_0\sim 1.52.10^{-80}$J$^{2}$.m$^{3}$ and $\sigma=0.13\,\mu$m \cite{Jen},
the disorder fraction inside the droplet is about $n_R/n_{\text{eq}} = 6\times10^{-4}$ ensuring the sufficient criterion for the weak disorder regime.

At zero temperature, the condensed depletion and the anomalous fraction can be calculated, respectively via Eqs.(\ref{nor}) and (\ref{anom}). This yields 
\begin{equation} \label{nor1}
\frac{\tilde{m}_{0\pm}}{n}=\frac{3\tilde{n}_{0\pm}}{n}=\frac{15}{96\sqrt{2}} \sqrt{\frac{n}{n^{(0)}}} \left\vert \frac{\delta a_+}{a} \right\vert \left(\frac{\delta a_{\pm}}{a}\right)^{3/2}.
\end{equation}
Here again the standard regularization scheme is introduced in the anomalous density in order to remove the well-known ultraviolet divergence problem caused
by the use of the contact interparticle interactions \cite{Boudj18}.

In the droplet phase, one should neglect the complex component ($\delta a_+/a \ll 1$) of the noncondensed and anomalous densities \cite{Boudj21}.
The profiles of the total depletion $\tilde n=\tilde n_{0-}+ n_R$ and the anomalous density $\tilde m=\tilde m_{0-}+ n_R$ are shown in Figs \ref{GL}.(b) and \ref{GL}.(c).
We observe  that the depletion and the anomalous correlation of the droplet decrease with $a_{12}/a$ and $\sigma/\xi$.
Remarkably, the anomalous density is somehow insensitive to the disorder correlation length (slightly decays with $\sigma/\xi$, see Fig.\ref{GL}.(c)).

\subsection{Finite temperature}\label{Ttf}

At finite temperature the properties of the droplet can be analyzed by minimizing the free energy which is defined as  \cite{Boudj18,Boudj9}:
\begin{align} \nonumber
F&=E+\frac{T}{2\pi^2} \int_0^{\infty} dk\, k^2 \bigg [\ln\left(\frac{2}{\sqrt{I_{k+}}+1}\right) \nonumber\\
&+  \ln\left(\frac{2}{\sqrt{I_{k-}}+1}\right) \bigg].\nonumber
\end{align}
In terms of the equilibrium density it turns out to be given by
\begin{equation}\label {fergy1}
\frac{F}{\vert E_{0}\vert}=\frac{E}{\vert E_{0}\vert}-\sum_{\pm}\frac{\sqrt{2}\pi^4(\delta a_+/a)^{4} (n/n^{(0)})^{-5/2}}{124416 (\delta a_{\pm}/a)^{3/2}}
\left(\frac{T}{\vert E_{0}\vert/N}\right)^{4},
\end{equation}
which is divergent when the density tends to zero.

\begin{figure}
\centerline{
\includegraphics[scale=0.82]{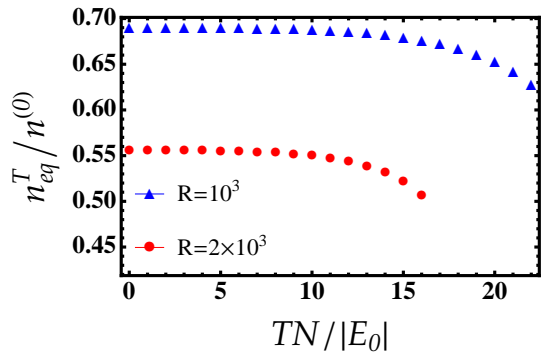}}
 \caption{ Thermal equilibrium density $n_{\text{eq}}^T/n^{(0)}$ as a function of the temperature for different values of the disorder strength and $a_{12}/a=-1.1$ and $\sigma/\xi=0.63$.}
\label{theq} 
\end{figure}

The  temperature dependence of the equilibrium density and the critical temperature of the droplet can be determined by setting $\partial F/\partial n=0$, 
and $\delta a_+/a \ll 1$ \cite {Petrov,Boudj18}.
Figure.\ref{theq} shows that the thermal equilibrium density $n_{\text{eq}}^T$ exhibits very weak temperature dependence at $T \lesssim  13|E_0|/N$,
while it reduces for higher temperatures. 
We see also that it sorely diminishes with the disorder strength, indicating that the droplet becomes strongly depleted.

\begin{figure}
\centerline{
\includegraphics[scale=0.8]{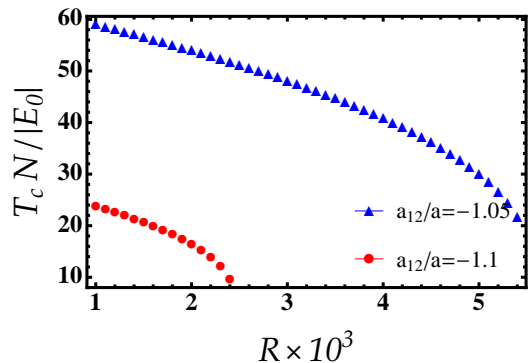}}
 \caption{ Critical temperature as a function of the  disorder strength $R$ for different values of $a_{12}/a$ and  $\sigma/\xi=0.63$.}
\label{CT} 
\end{figure}

Effects of the disorder strength on the critical temperature are displayed in Fig.\ref{CT}.
We see that $T_c$ decreases with increasing $R$ and with $|a_{12}/a|$.

\section{Generalized disorder Gross-Pitaevskii equation} \label{GDGPE}

To better understand effects of disorder on the droplet state, we will derive in this section the generalized disorder
GPE and solve it numerically.

In the miscible phase and close to the collapse point, we can describe the system with an effective low-energy
theory, an effective single component GPE, and we consider identical spatial modes for the two components \cite{Petrov}.
We then will introduce two assumptions: 
(i) The condensate, the thermal cloud and the anomalous correlation must vary slowly at the scale of the extended healing length \cite{Boudj18,Boudj21}. 
(ii)  The disorder potential changes smoothly in space on a length scale comparable to the healing length \cite{Boudj44,Nagler}.

For the sake of simplicity, we neglect the fluctuations induced by disorder potential. 
The functional  energy associated with the equation of state (\ref{genergy1}) which describes a self-bound droplet 
subjected to an external disorder, can be written in the following dimensionless form:
\begin{equation}
{\cal E} (\phi,\phi^*)=\frac{1}{2}\vert\nabla\phi\vert^{2}-\frac{3}{2}\vert\phi\vert^{4}+\sqrt{\frac{n_{\text{eq}}}{n^{(0)}}}\vert\phi\vert^{5}+\tilde{U}\vert\phi\vert^{2},
\end{equation}
where $\tilde{U} =U \tau/\hbar$, and $\tau=6\hbar/n_{\text{eq}}g\vert \delta a/a_{+}\vert$.
The corresponding GPE can be derived using $i \partial \phi /\partial \tilde{t} =\partial {\cal E}/\partial \phi^*$. This yields
\begin{align} \label{DisGPE}
i\frac{\partial\phi(\tilde{\textbf{r}},\tilde{t})}{\partial \tilde{t}}=\left(-\frac{1}{2}\Delta_{\tilde{\textbf{r}}}-3\vert\phi\vert^{2}+\frac{5}{2}\sqrt{\frac{n_{\text{eq}}}{n^{(0)}}}\vert\phi\vert^{3}
+\tilde{U}(\tilde{\textbf{r}})\right)\phi(\tilde{\textbf{r}},\tilde{t}),
\end{align}
where we introduced the rescaled coordinate $\tilde{r}= r\sqrt{m/\hbar \tau}$ and the rescaled time $ \tilde{t}=t/\tau$.
Equation (\ref{DisGPE}) can be thought of as describing self-consistently  the LHY and disorder effects.
The stationary solutions can be found via the transformation: $\phi(\tilde{\textbf{r}},\tilde{t})= \phi(\tilde{\textbf{r}}) \exp( {-i\mu \tilde{t}})$.

We consider now a Gaussian correlated disorder potential defined as \cite{Adh}:
\begin{equation}
\tilde U (\tilde{\textbf{r}})=\tilde U_{0}\sum_{j=1}^{M}f(\tilde{\textbf{r}}-\tilde{\textbf{r}}_{j}),
\end{equation}
where $M$ is the number of impurities, $U_{0}$ is the amplitude, and ${\tilde r}_{j}$ are the uncorrelated random positions, 
and $f$ is a real-valued function of width $\sigma$ and has Gaussian-shaped impurities $ f(\tilde{\textbf{r}})= e^{-\tilde{\textbf{r}}^{2}/\tilde{\sigma}^{2}}$
with $\tilde{\sigma}= \sigma\sqrt{m/\hbar \tau}$ being dimensionless the characteristic length of the disorder.

The ground-state density profile of the disordered droplet is obtained from the numerical integration of the stationary GPE (\ref{DisGPE})
using the split-step method, which is based on the fast Fourier transforms \cite{Sem}. 
To generate the Gaussian potential we use a set of random numbers which are then mapped into the interval $[-L,L]$ by a linear transformation.
We choose $M=300$, $L=30$, and a small width \cite{Adh}.

\begin{figure}
\centerline{
\includegraphics[scale=0.7] {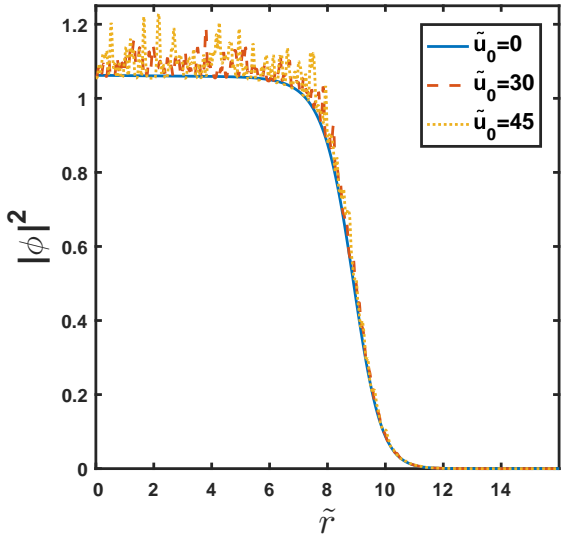}}
 \caption{ Density profiles of the droplet for different values of the disorder strength for $N=3000$, $\tilde\sigma=0.1$, and $a_{12}/a=-1.05$.}
\label{DP}
\end{figure}

In Fig.\ref{DP} we plot the density profiles as a function of the radial distance.
Two fascinating properties are observed here. First, the density in the plateau region increases and varies fastly with the disorder strength.
Second, the situation is completely different at the edge of the droplet,  where the atomic density varies slowly even for large disorder strength since
the LHY and interaction energies dominate  over the disorder effects.

\section{Collective modes} \label{modes}

This section deals with the collective modes of disordered droplets. A useful way to qualitatively  or even quantitatively analyze  them is to use a variational method.
We then consider a simplified Gaussian variational ansatz: 
\begin{equation} \label{Gauss}
\phi(\tilde{r},\tilde{t})=\sqrt{\frac{\tilde{N}}{\pi^{3/2} q(\tilde{t})^{3}}} \exp\left[\frac{-\tilde{r}^{2}}{2q(\tilde{t})^{2}}+i\gamma(\tilde{t})\tilde{r}^{2}\right],
\end{equation}
where the variational parameters are the droplet width $q$,  and the phase $\gamma$.
The normalization factor ensures the conservation of the condition: $\int d^3 \tilde{r} \phi(\tilde{r},\tilde{t}) = \tilde{N}$, where
$\tilde{N} =\left(m/\hbar\tau\right)^{3/2} N/n_{\text{eq}}$, with $N$ being the total number of particles in the droplet.
We replace the ansatz (\ref{Gauss}) in the density Lagrangian ${\cal L}=(i/2)\left[\phi\, (d\phi^{*}/dt)-\phi^{*} (d\phi/dt)\right]+ {\cal E}$, 
and obtain the Lagrangian $L= \int_0^{\infty} d^3 \tilde{r}  {\cal L} $:
\begin{align}
\frac{L}{\tilde{N}}&=\frac{3}{2}\dot{\gamma}q^{2}+3\gamma^{2}q^{2}+\frac{3}{4q^{2}}-\frac{3\tilde{N}}{2^{5/2}\pi^{3/2}q^{3}}
+\sqrt{\frac{n_{\text{eq}}}{n^{(0)}}}\frac{2^{3/2}\tilde{N}^{3/2}}{\pi^{9/4}5^{3/2}q^{9/2}} \nonumber\\
&+\sum_{j=1}^{M}\tilde{U}_{0}\frac{\tilde{\sigma}^{3}}{(q^{2}+\tilde{\sigma}^{2})^{3/2}}\exp\left[\frac{-\tilde{r}_{j}^{2}}{\tilde{\sigma}^{2}}\left(1-\frac{q^{2}}{q^{2}+\tilde{\sigma}^{2}}\right)\right].
\end{align}
The corresponding Euler-Lagrange equations of motion read:
\begin{equation}
\gamma=\frac{1}{2q} \frac{dq}{dt},
\end{equation}
and 
\begin{equation}
\frac{d^{2}q}{dt^{2}}=-\frac{dU_{\text{eff}}(q)}{dq},
\end{equation}
where we have introduced the effective potential for oscillations of the droplet width
\begin{align}
U_{\text{eff}}(q)&=\frac{1}{2q^{2}}-\frac{\tilde{N}}{2^{3/2}\pi^{3/2}q^{3}}+\sqrt{\frac{n_{eq}}{n^{(0)}}}\frac{2^{5/2}\tilde{N}^{3/2}}{3\pi^{9/4}5^{3/2}q^{9/2}} \\
&+\frac{2}{3}\sum_{j=1}^{M}\tilde{U}_{0}\frac{\tilde{\sigma}^{3}}{(q^{2}+\tilde{\sigma}^{2})^{3/2}} \exp\left[\frac{-\tilde{r}_{j}^{2}}{\tilde{\sigma}^{2}}\left(1-\frac{q^{2}}{q^{2}+\tilde{\sigma}^{2}}\right)\right]. \nonumber
\end{align}

\begin{figure}
\includegraphics[scale=0.72]{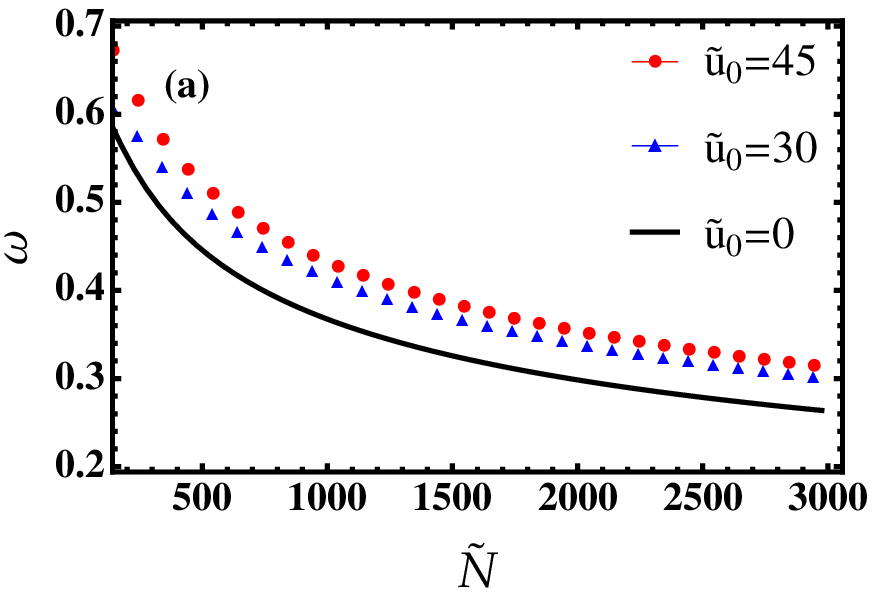}
\includegraphics[scale=0.7]{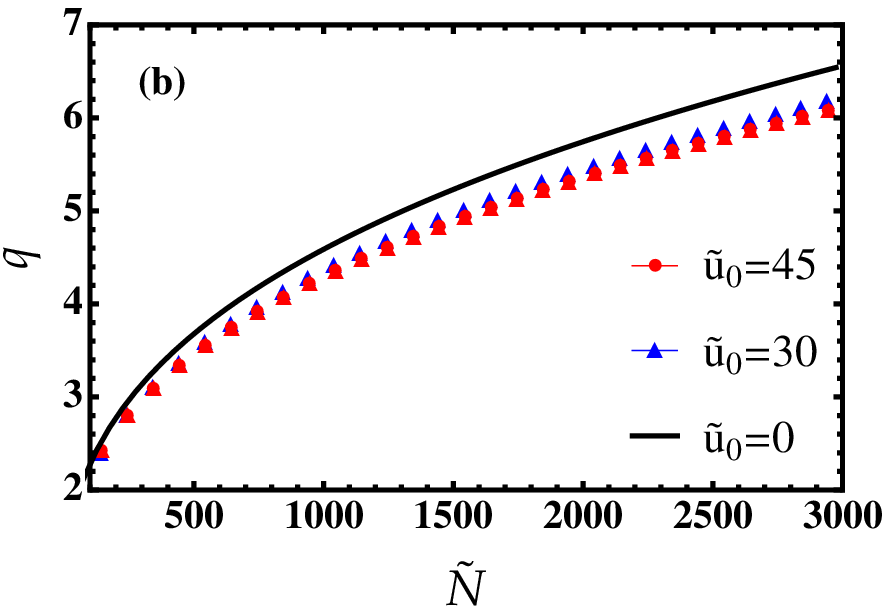}
 \caption{ (a) Excitation frequencies in units of $1/\tau$ as a function of the particle number for several values of $\tilde U_{0}$.
(b) Droplet width as a function of the particle number for several values of $\tilde U_{0}$. Parameters are $a_{12}/a=-1.05$ and $\tilde \sigma=0.1$.}
\label{BrMod}
\end{figure}
The low-lying excitations around the equilibrium solutions $q_{0}$ (droplet minimum) can be computed by using the linearization $q(t)=q_{0}+\delta q(t)$, where $ \delta q(t)\ll q_{0}$,
and $\delta q(t)=\delta q e^{i\omega t}$. This gives for the frequencies of the breathing modes: 
\begin{equation} \label{BMod}
\omega^{2}=\frac{d^{2}U_{\text{eff}}(q)}{dq^{2}}\Big\vert_{q=q_{0}}.
\end{equation}

In Fig.\ref{BrMod} we represent the resulting oscillation frequencies as predicted by Eq.(\ref{BMod}).
It is interesting to observe that the frequencies of the breathing modes of the droplet are increasing with the disorder strength
in the whole range of atoms number as shown in Fig.\ref{BrMod}.a.

Figure \ref{BrMod}.b depicts that the droplet width $q$ increases with the particle number while it decreases with the disorder strength notably for large $\tilde N$.

\section{Conclusions}\label{Conc}

In this paper, we studied the effects of a weak random potential with a Gaussian correlation function on the properties of self-bound droplets.
Within the Bogoliubov approach, we calculated corrections due to the disorder fluctuations to the energy and to the condensed density. 
The equilibrium density, the quantum depletion, and the anomalous density have also been obtained in terms of the disorder parameters and the interspecies interactions.
We showed that the disorder decreases the energy leading to destabilize the droplet state.    
The critical disorder strength above which the droplet evaporates has been accurately established.
One can expect that for strong disorder, the self-bound droplet breaks down to form several mini droplets.
At finite temperature, we found that the disorder plays a crucial role in the thermal density equilibrium and in the critical temperature of the self-bound liquid.
Additionally, we derived self-consistently a generalized disorder GPE and solve it numerically using a suitable scheme.
Our results indicate that the density follows the modulation of the disorder in bulk and becomes important for large  disorder strength.
We also  examined the influence of the disorder on the width and the breathing modes of droplets via a Gaussian variational ansatz.

It is clear that the results we predict in this work are achievable in current experimental setups.
Our study not only bridges the gap between ultradilute droplets and disorder but also elucidates the localization phenomenon of droplets in binary BECs.

\section{Acknowledgments}
We thank Axel Pelster  for valuable comments on the manuscript.
We acknowledge support from the Algerian Ministry of Higher Education and Scientific Research under Research Grant No. PRFU-B00L02UN020120190001.

\end{document}